\documentclass{article}

\usepackage{amsmath}
\usepackage{natbib}
\usepackage[title]{appendix}
\usepackage[margin=1.0in]{geometry}
\usepackage{times}
\usepackage[symbol]{footmisc}
\usepackage{enumitem}
\allowdisplaybreaks

\numberwithin{equation}{section}

\providecommand{\keywords}[1]
{
  \hangindent=1.937cm
  {\small{\text{\textit{Keywords:}} #1}}
}

\title{{\bf Addendum: Hidden symmetries, trivial conservation laws and Casimir invariants in geophysical fluid dynamics} (2018 {\it J.\ Phys.\ Commun.} {\bf 2} 115018)}

\author{Martin Charron and Ayrton Zadra \\
	Recherche en pr\'evision num\'erique atmosph\'erique \\
	Environnement et Changement climatique Canada, Dorval, Qc, Canada}

\date{\today}

\begin{document}

\maketitle

\begin{abstract}
An extension is proposed to the internal symmetry transformations associated with mass, entropy and other Clebsch-related conservation in geophysical fluid dynamics. Those symmetry transformations were previously parameterized with an arbitrary function $\cal F$ of materially conserved Clebsch potentials. The extension consists in adding potential vorticity $q$ to the list of fields on which a new arbitrary function $\cal G$ depends. If ${\cal G}=q{\cal A}(s)$, where ${\cal A}(s)$ is an arbitrary function of specific entropy $s$, then the symmetry is trivial and gives rise to a trivial conservation law. Otherwise, the symmetry is non-trivial and an associated non-trivial conservation law exists. Moreover, the notions of trivial and non-trivial Casimir invariants are defined. All non-trivial symmetries that become hidden following a reduction of phase space are associated with non-trivial Casimir invariants of a non-canonical Hamiltonian formulation for fluids, while all trivial conservation laws are associated with trivial Casimir invariants.
\end{abstract}

\keywords{potential vorticity; global gauge symmetry; trivial and non-trivial Casimir invariants; phase-space reduction.}

\noindent\rule{\textwidth}{0.3pt}

\section{Introduction}
An explicit association between, on the one hand, hidden global gauge symmetries and trivial conservation laws and, on the other hand, Casimir invariants in geophysical fluid dynamics was described in \citet{Charron18b}. However, the symmetry transformations proposed in \citet{Webb17,Charron18b} are not as general as they may be, insofar as they do not involve a dependence on potential vorticity, which restricts the interpretation of a certain class of Casimir invariants as resulting from hidden symmetries. The internal symmetry presented in \citet{Charron18b} is here generalized, allowing a complete characterization of known Casimir invariants in geophysical fluid dynamics. The generalization consists in symmetry transformations involving an arbitrary function of potential vorticity. In this context, it is useful to define the notions of trivial and non-trivial Casimir invariants. As will be seen below, the former are associated with trivial conservation laws and the latter with non-trivial hidden symmetries. The following results are compatible with the fact that the conservation law for potential vorticity density is trivial, however it turns out that the conservation law for the density of an otherwise arbitrary non-linear function of potential vorticity is non-trivial.

In section \ref{defsym}, internal symmetry transformations of Clebsch fields involving potential vorticity are presented and their associated conserved 4-current is obtained. In section \ref{trin}, the notions of trivial and non-trivial Casimir invariants are defined. A discussion is presented in section \ref{disc} in which it is shown that all Casimir invariants in geophysical fluid dynamics may be characterized either as trivial---if they are associated with trivial conservation laws---or as non-trivial---if they are associated with non-trivial hidden symmetries.

\section{An internal symmetry involving potential vorticity}\label{defsym}
It was shown in \cite{Charron18b} that the Lagrangian density $\cal L$, based on Clebsch potentials and suitable for geophysical fluid dynamics, is invariant under certain internal transformations of dynamical fields also called global gauge transformations (see their subsection 3.4), giving rise via Noether's first theorem to mass, entropy and other Clebsch-related conservation. Those symmetry transformations were parameterized with an arbitrary and sufficiently smooth function ${\cal F}={\cal F}(s,\gamma_{(1)},\lambda_{(1)},...,\gamma_{(N)},\lambda_{(N)})$\footnote[1]{Definitions and the notation are identical to those of \cite{Charron18b}. The reader is invited to first consult that paper. The expression ``sufficiently smooth'' refers to the requirement that the field transformations be single-valued and well-defined at any space-time point.}, which depends on specific entropy and other Clebsch pairs (but not on their space-time derivatives), all materially conserved on-shell.

That symmetry is here generalized by adding potential vorticity $q$ to the list of fields on which the arbitrary function depends, i.e.\ by replacing ${\cal F}$ with an arbitrary and sufficiently smooth function ${\cal G}={\cal G}(q,s,\gamma_{(1)},\lambda_{(1)},...,\gamma_{(N)},\lambda_{(N)})$. The following internal transformations of dynamical fields:
\begin{subequations}
\begin{align}
\bar\delta\rho &= 0, \label{s1} \\
\bar\delta\alpha &= \epsilon \left( {\cal G}-q\frac{\partial{\cal G}}{\partial q} -\sum_{r=1}^N\gamma_{(r)}\left\{\frac{\partial{\cal G}}{\partial \gamma_{(r)}} - \left[\frac{\partial{\cal G}}{\partial q}\right]_{,\mu} \frac{\varepsilon^{0\mu\sigma\nu}}{\sqrt{g}\rho} s_{,\nu} \lambda_{(r),\sigma}\right\} \right), \\
\bar\delta\beta &= -\epsilon \left( \frac{\partial{\cal G}}{\partial s} + \left[\frac{\partial{\cal G}}{\partial q}\right]_{,\mu}\sum_{r=1}^N \frac{\varepsilon^{0\mu\sigma\nu}}{\sqrt{g}\rho} \gamma_{(r),\nu} \lambda_{(r),\sigma} \right), \\
\bar\delta s &= 0, \label{ds} \\
\bar\delta\gamma_{(r)} &= -\epsilon \left( \frac{\partial{\cal G}}{\partial \lambda_{(r)}} +\left[\frac{\partial{\cal G}}{\partial q}\right]_{,\mu} \frac{\varepsilon^{0\mu\sigma\nu}}{\sqrt{g}\rho} s_{,\nu}\gamma_{(r),\sigma} \right), \qquad (r=1,...,N), \\
\bar\delta\lambda_{(r)} &= \epsilon \left( \frac{\partial{\cal G}}{\partial \gamma_{(r)}} -\left[\frac{\partial{\cal G}}{\partial q}\right]_{,\mu} \frac{\varepsilon^{0\mu\sigma\nu}}{\sqrt{g}\rho} s_{,\nu}\lambda_{(r),\sigma} \right), \qquad (r=1,...,N), \label{s2}
\end{align}
\end{subequations}
where $\epsilon$ is an infinitesimal constant, are symmetry transformations of the dynamics\footnote[2]{Notice that gauge freedom cannot exist for measurable physical quantities. Therefore, the density $\rho$ and specific entropy $s$, being measurable fields, cannot change under gauge transformations. This explains \eqref{s1} and \eqref{ds}.}. This may be seen by establishing how the Lagrangian density $\cal L$, given by (2.5) in \cite{Charron18b}, changes off-shell under these transformations. It turns out that $\bar\delta{\cal L}={J^\mu}_{:\mu}$ to order $\epsilon$, where
\begin{align}
J^\mu = \epsilon \rho\frac{\partial{\cal G}}{\partial q} \left[g^{0\mu}q-\sum_{r=1}^N \frac{\varepsilon^{0\mu\sigma\nu}}{\sqrt{g}\rho} \left( s^{:0}\gamma_{(r),\sigma} \lambda_{(r),\nu} + s_{,\nu} {\gamma_{(r)}}^{:0}\lambda_{(r),\sigma} + s_{,\sigma}\gamma_{(r),\nu} {\lambda_{(r)}}^{:0} \right) \right]
\end{align}
(Appendix \ref{demo}, demonstration 1). Because the Lagrangian density is invariant up to a divergence under the infinitesimal transformations \eqref{s1}--\eqref{s2}, they represent a symmetry of the dynamics. The associated conserved 4-current $k^\mu$ is given by
\begin{align}
k^\mu &= J^\mu - \sum_{p=1}^P \frac{\partial{\cal L}}{\partial (\psi_{(p),\mu})} \bar\delta\psi_{(p)}=J^\mu+\rho u^\mu \left( \bar\delta\alpha+\sum_{r=1}^N \gamma_{(r)}\bar\delta\lambda_{(r)} \right), \nonumber \\
&=\epsilon\rho u^\mu {\cal G} + J^\mu-\epsilon\rho u^\mu q\frac{\partial{\cal G}}{\partial q}.
\end{align}
It may be verified that $J^\mu=\epsilon\rho u^\mu q\,\partial{\cal G}/\partial q$ on-shell (Appendix \ref{demo}, demonstration 2), and therefore the conserved 4-current associated with the symmetry transformations \eqref{s1}--\eqref{s2} is $k^\mu=\epsilon\rho u^\mu{\cal G}$ on-shell.

\section{Trivial and non-trivial Casimir invariants}\label{trin}
Casimir invariants are distinguished functionals $C$ \citep{Olver93} which are constants of the motion and which satisfy the equations
\begin{align}
\sum_{b=1}^P\int_{\cal D} d^3y \,{\cal J}_{(ab)}({\mathbf x},{\mathbf y})\frac{\delta C}{\delta\eta_{(b)}({\mathbf y})}=0 \quad\text{for } a=1,...,P \text{ and for all $\mathbf x$ in the interior of $\cal D$,}
\end{align}
where $\eta_{(b)}$ is a dynamical field and ${\cal J}_{(ab)}({\mathbf x},{\mathbf y}) \equiv [\eta_{(a)}({\mathbf x}),\eta_{(b)}({\mathbf y})]$ a Hamiltonian operator. These equations are trivially satisfied if the first variation $\delta C$ of the functional $C$ vanishes on-shell. All charges $C$ obtained from trivial conservation laws satisfy $\delta C=0$ on-shell. Trivial conservation laws of the first kind, for which the integrand of $C$ vanishes on-shell \citep{Olver93}, have vanishing first variations. Trivial conservation laws of the second kind, for which
\begin{align}
C=\int_{\cal D} d^3x\,\left(\sqrt{g}F^{0i}\right)_{,i}=\oint_{\partial{\cal D}} dS_i \,F^{0i} \quad\text{with } F^{\mu\nu}=-F^{\nu\mu},
\end{align}
also satisfy $\delta C=0$---because, in this latter case, $C$ is a boundary term. These Casimir invariants will be called {\it trivial}.

If the kernel of the Hamiltonian operator ${\cal J}$ is not empty, Casimir invariants with $\delta C \ne 0$, and whose functional derivatives $\delta C/\delta\eta$ therefore lie in the phase subspace defined by the kernel of ${\cal J}$, will be called {\it non-trivial}.

\section{Discussion}\label{disc}
The right-hand sides of the symmetry transformations \eqref{s1}--\eqref{s2} vanish in the special case ${\cal G}=q{\cal A}(s)$, where ${\cal A}(s)$ is an arbitrary function of specific entropy $s$. The symmetry is then said to be trivial \citep{Olver93}, which is consistent with the fact that the conservation law for potential vorticity density is trivial (i.e.\ the special case ${\cal A}=\text{constant}$), as shown in \citet{Charron18b, Charron18}. The conserved on-shell 4-current $k^\mu=J^\mu=\epsilon \rho u^\mu q{\cal A}(s)$ is therefore associated with null (trivial) transformations. All other choices for ${\cal G}={\cal G}(q,s,\gamma_{(1)},\lambda_{(1)},...,\gamma_{(N)},\lambda_{(N)})$ result in non-trivial symmetries and non-trivial conservation laws when $N \ge 2$\footnote[3]{The case $N < 2$ does not allow to fully represent certain flow configurations with Clebsch fields and is not considered here.}. Consequently, the conservation law for the density of an arbitrary local function ${\cal G}={\cal Q}(q)$ of potential vorticity $q$ is non-trivial and associated with a symmetry via Noether's first theorem, unless this arbitrary function is $bq$, where $b$ is any constant. This symmetry is a global internal---or global gauge---symmetry of the dynamics in arbitrary coordinates, and is unrelated to particle relabeling, which is a coordinate transformation not resulting in a dynamically relevant symmetry, as shown from (3.22), (3.23), (4.16) and (4.17) in \citet{Charron18}.

The non-canonical Hamiltonian operator resulting from all the Dirac brackets between dynamical fields from the extended phase space $(t,\pi_{(t)},\rho,\alpha,\beta,s,\gamma_{(1)},\lambda_{(1)},...,\gamma_{(N)},\lambda_{(N)})$ is, for $N=2$,
\begin{align}
{\cal J}({\mathbf x},{\mathbf y})=\frac{\delta^{(3)}({\mathbf x}-{\mathbf y})}{\sqrt{g}\rho}\left(
\begin{array}{cccccccccc}
 0 & \sqrt{g}\rho & 0 & 0 & 0 & 0 & 0 & 0 & 0 & 0 \\
-\sqrt{g}\rho & 0 & \rho^2 (\sqrt{g})_{,0}  & 0 & 0 & 0 & 0 & 0 & 0 & 0 \\
 0 & -\rho^2 (\sqrt{g})_{,0} & 0 & \rho & 0 & 0 & 0 & 0 & 0 & 0 \\
 0 & 0 & -\rho & 0 & \beta & 0 & \gamma_{(1)} & 0 & \gamma_{(2)} & 0 \\
 0 & 0 & 0 & -\beta & 0 & 1 & 0 & 0 & 0 & 0 \\
 0 & 0 & 0 & 0 & -1 & 0 & 0 & 0 & 0 & 0 \\
 0 & 0 & 0 & -\gamma_{(1)} & 0 & 0 & 0 & 1 & 0 & 0 \\
 0 & 0 & 0 & 0 & 0 & 0 & -1 & 0 & 0 & 0 \\
 0 & 0 & 0 & -\gamma_{(2)} & 0 & 0 & 0 & 0 & 0 & 1 \\
 0 & 0 & 0 & 0 & 0 & 0 & 0 & 0 & -1 & 0
\end{array}
\right).
\end{align}
It may be verified that the determinant of this Hamiltonian operator scales as $\rho^2/(\sqrt{g}\rho)^{4+2N}$ and is non-vanishing. This means that this Hamiltonian operator does not admit non-trivial Casimir invariants. In this example, because all non-trivial conservation laws are explicit and not hidden, the only Casimir invariants must be trivial. The volume integral of potential vorticity density provided by (5.70) in \citet{Charron18b}, being a boundary term, is such a trivial Casimir invariant. More generally, the integral over space of $\sqrt{g}\rho q{\cal A}(s)$ is a trivial Casimir invariant (Appendix \ref{demo}, demonstration 3). Other trivial Casimir invariants are easily constructed by defining an antisymmetric tensor, for instance $\rho^{:\mu}s^{:\nu}-\rho^{:\nu}s^{:\mu}$, and a 4-current as $c^\mu\equiv (\rho^{:\mu}s^{:\nu}-\rho^{:\nu}s^{:\mu})_{:\nu}$. The integral over space of $\sqrt{g}\,c^0$ is necessarily a trivial Casimir invariant, being a boundary term.

From the relation $v_i=u_i=\alpha_{,i}+\beta s_{,i} +\sum_{r=1}^N \gamma_{(r)}\lambda_{(r),i}$, it may be verified that \eqref{s1}--\eqref{s2} lead to $\bar\delta u_i=0$ (Appendix \ref{demo}, demonstration 1) and $\bar\delta u^i=0$ (this from $u^i=h^{ij}u_j+g^{0i}$) to order $\epsilon$. Therefore, the Hamiltonian functional $H[t,\pi_{(t)},\rho,\alpha,\beta,s,\gamma_{(1)},\lambda_{(1)},...,\gamma_{(N)},\lambda_{(N)}]$ provided by (5.39) in \citet{Charron18b} may become a functional of the fields $[t,\pi_{(t)},\rho,s,u^1,u^2,u^3]$---(5.50) in \citet{Charron18b}---as a result of the phase-space reduction which follows from the symmetry \eqref{s1}--\eqref{s2}. The corresponding Hamiltonian operator in the reduced phase space, which necessarily depends (possibly non-locally) on $(t,\pi_{(t)},\rho,s,u^1,u^2,u^3)$ alone, is obtained from (5.46), (5.52)--(5.58) in \citet{Charron18b}. It may be verified that its determinant vanishes\footnote[4]{This may be seen intuitively by discretizing the Hamiltonian operator in the reduced phase space and finding its determinant in finite dimensions, which is zero. The determinant remains zero in the infinite-dimensional limit.} and therefore this Hamiltonian operator in the reduced phase space admits non-trivial Casimir invariants. They are all associated via Noether's first theorem with the conservation laws resulting from the symmetry \eqref{s1}--\eqref{s2}---which becomes hidden after the phase-space reduction---and are provided by
\begin{align}
C=\int_{\cal D} d^3x\sqrt{g}\rho\,{\cal C}_1(q,s). \label{cr}
\end{align}
In this case, the arbitrary ${\cal C}_1(q,s)$ is a function of the remaining dynamical fields after the reduction of the phase space, i.e.\ the explicit dependence of $\cal G$ on the Clebsch pairs $\gamma_{(r)},\lambda_{(r)}$ is dropped and $\cal G$ is then written as ${\cal C}_1(q,s)$. The conserved charge and Casimir invariant \eqref{cr} could not be completely associated with a hidden symmetry in \citet{Charron18b} because the considered symmetry transformations (3.30)--(3.35) in that paper were not sufficiently general. It has been shown here that {\it all} non-trivial Casimir invariants in the reduced phase space are associated with a well-identified, non-trivial but hidden, global gauge symmetry---even those Casimir invariants involving an arbitrary function of potential vorticity $q$, unless this function is ${\cal C}_1=q{\cal A}(s)$ in which case the symmetry is trivial.

\section*{Acknowledgements} The authors thank Peter Olver for an exchange on trivial Casimir invariants. They also thank Christopher Subich and St\'ephane Gaudreault for their comments on an earlier draft of this addendum.

\begin{appendices}
\section{Demonstrations}\label{demo}
Consider the quantity ${\cal G}' \equiv \epsilon\,{\cal G}(q,s,\gamma_{(1)},\lambda_{(1)},...,\gamma_{(N)},\lambda_{(N)})$ and the definition $u^\mu \equiv h^{\mu\nu}v_\nu+g^{0\mu}$, where
\begin{align}
q \equiv \frac{\varepsilon^{0\mu\sigma\nu}}{\sqrt{g}\rho} u_{\sigma:\mu} s_{,\nu}=\frac{\varepsilon^{0ijk}}{\sqrt{g}\rho} u_{j,i} s_{,k} = \frac{\varepsilon^{0ijk}}{\sqrt{g}\rho} v_{j,i} s_{,k}
\end{align}
with
\begin{align}
u_j=v_j \equiv \alpha_{,j}+\beta s_{,j}+\sum_{r=1}^N \gamma_{(r)}\lambda_{(r),j},
\end{align}
but $u_0 \ne v_0$. Therefore,
\begin{align}
q=\sum_{r=1}^N\frac{\varepsilon^{0\mu\sigma\nu}}{\sqrt{g}\rho} \gamma_{(r),\mu}\lambda_{(r),\sigma} s_{,\nu} \quad\text{off-shell.}
\end{align}
\begin{enumerate}[leftmargin=*]
\item From the definition
\begin{align}
v_\tau \equiv \alpha_{,\tau}+\beta s_{,\tau}+\sum_{r=1}^N \gamma_{(r)}\lambda_{(r),\tau}
\end{align}
and the field transformations \eqref{s1}--\eqref{s2}, the vector field perturbation $\bar\delta v_\tau$ is (to first order in $\epsilon$)
\begin{align}
\bar\delta v_\tau &= (\bar\delta\alpha)_{,\tau}+s_{,\tau} \bar\delta\beta+\beta(\bar\delta s)_{,\tau} + \sum_{r=1}^N \left[ \lambda_{(r),\tau} \bar\delta\gamma_{(r)} + \gamma_{(r)}(\bar\delta\lambda_{(r)})_{,\tau}\right], \nonumber \\
&= (\bar\delta\alpha)_{,\tau}+s_{,\tau} \bar\delta\beta+ \sum_{r=1}^N \left[ \lambda_{(r),\tau} \bar\delta\gamma_{(r)} + \gamma_{(r)}(\bar\delta\lambda_{(r)})_{,\tau}\right], \nonumber \\
&= {\cal G}'_{,\tau}-q_{,\tau} \frac{\partial{\cal G}'}{\partial q}-q\left( \frac{\partial{\cal G}'}{\partial q} \right)_{,\tau} - \sum_{r=1}^N \gamma_{(r),\tau} \frac{\partial{\cal G}'}{\partial \gamma_{(r)}}-\sum_{r=1}^N \gamma_{(r)} \left(\frac{\partial{\cal G}'}{\partial \gamma_{(r)}}\right)_{,\tau} \nonumber \\
&\quad\, + \left(\frac{\partial{\cal G}'}{\partial q}\right)_{,\mu:\tau} \sum_{r=1}^N \frac{\varepsilon^{0\mu\sigma\nu}}{\sqrt{g}\rho} s_{,\nu} \gamma_{(r)}\lambda_{(r),\sigma} +\left(\frac{\partial{\cal G}'}{\partial q}\right)_{,\mu} \sum_{r=1}^N \frac{\varepsilon^{0\mu\sigma\nu}}{\sqrt{g}} \left(\frac{1}{\rho}s_{,\nu} \gamma_{(r)}\lambda_{(r),\sigma}\right)_{:\tau}-s_{,\tau} \frac{\partial{\cal G}'}{\partial s} \nonumber \\
&\quad\, - \left(\frac{\partial{\cal G}'}{\partial q}\right)_{,\mu}\sum_{r=1}^N \frac{\varepsilon^{0\mu\sigma\nu}}{\sqrt{g}\rho} s_{,\tau}\gamma_{(r),\nu} \lambda_{(r),\sigma} - \sum_{r=1}^N \lambda_{(r),\tau}\frac{\partial{\cal G}'}{\partial \lambda_{(r)}} - \left(\frac{\partial{\cal G}'}{\partial q}\right)_{,\mu} \sum_{r=1}^N \frac{\varepsilon^{0\mu\sigma\nu}}{\sqrt{g}\rho} s_{,\nu}\gamma_{(r),\sigma} \lambda_{(r),\tau} \nonumber \\
&\quad\, + \sum_{r=1}^N \gamma_{(r)} \left(\frac{\partial{\cal G}'}{\partial \gamma_{(r)}}\right)_{,\tau} - \left(\frac{\partial{\cal G}'}{\partial q}\right)_{,\mu:\tau} \sum_{r=1}^N \frac{\varepsilon^{0\mu\sigma\nu}}{\sqrt{g}\rho} s_{,\nu} \gamma_{(r)}\lambda_{(r),\sigma} - \left(\frac{\partial{\cal G}'}{\partial q}\right)_{,\mu} \sum_{r=1}^N \frac{\varepsilon^{0\mu\sigma\nu}}{\sqrt{g}} \left(\frac{1}{\rho}s_{,\nu} \lambda_{(r),\sigma}\right)_{:\tau} \gamma_{(r)}, \nonumber \\
&=-q\left( \frac{\partial{\cal G}'}{\partial q} \right)_{,\tau} 
+\left(\frac{\partial{\cal G}'}{\partial q}\right)_{,\mu} \sum_{r=1}^N \frac{\varepsilon^{0\mu\sigma\nu}}{\sqrt{g}} \left(\frac{1}{\rho}s_{,\nu} \gamma_{(r)}\lambda_{(r),\sigma}\right)_{:\tau} - \left(\frac{\partial{\cal G}'}{\partial q}\right)_{,\mu}\sum_{r=1}^N \frac{\varepsilon^{0\mu\sigma\nu}}{\sqrt{g}\rho} s_{,\tau}\gamma_{(r),\nu} \lambda_{(r),\sigma} \nonumber \\
&\quad\, - \left(\frac{\partial{\cal G}'}{\partial q}\right)_{,\mu} \sum_{r=1}^N \frac{\varepsilon^{0\mu\sigma\nu}}{\sqrt{g}\rho} s_{,\nu}\gamma_{(r),\sigma} \lambda_{(r),\tau}
- \left(\frac{\partial{\cal G}'}{\partial q}\right)_{,\mu} \sum_{r=1}^N \frac{\varepsilon^{0\mu\sigma\nu}}{\sqrt{g}} \left(\frac{1}{\rho}s_{,\nu} \lambda_{(r),\sigma}\right)_{:\tau} \gamma_{(r)}, \nonumber \\
&=-q\left( \frac{\partial{\cal G}'}{\partial q} \right)_{,\tau}+\left(\frac{\partial{\cal G}'}{\partial q}\right)_{,\mu} \sum_{r=1}^N \frac{\varepsilon^{0\mu\sigma\nu}}{\sqrt{g}\rho} \left( s_{,\nu} \gamma_{(r),\tau}\lambda_{(r),\sigma} - s_{,\tau}\gamma_{(r),\nu} \lambda_{(r),\sigma} - s_{,\nu}\gamma_{(r),\sigma} \lambda_{(r),\tau} \right), \nonumber \\
&=-q\left[ \frac{\partial{\cal G}'}{\partial q} \right]_{,\tau}+\left[\frac{\partial{\cal G}'}{\partial q}\right]_{,\mu} \sum_{r=1}^N \frac{\varepsilon^{0\mu\sigma\nu}}{\sqrt{g}\rho} \left( s_{,\tau}\gamma_{(r),\sigma} \lambda_{(r),\nu} + s_{,\nu} \gamma_{(r),\tau}\lambda_{(r),\sigma} + s_{,\sigma}\gamma_{(r),\nu} \lambda_{(r),\tau} \right).
\end{align}
The identity
\begin{align}
\varepsilon^{0\mu\sigma\nu} A_\mu \left( B_i C_\sigma D_\nu + B_\nu C_i D_\sigma + B_\sigma C_\nu D_i \right) \equiv \varepsilon^{0\mu\sigma\nu} A_i B_\mu C_\sigma D_\nu
\end{align}
implies that $\bar\delta v_i=0$ (note however that $\bar\delta v_0 \ne 0$). The term $\rho\,\bar\delta v_\tau$ is
\begin{align}
\rho\,\bar\delta v_\tau &= -\rho q\left[ \frac{\partial{\cal G}'}{\partial q} \right]_{,\tau}+\left[\frac{\partial{\cal G}'}{\partial q}\right]_{,\mu} \sum_{r=1}^N \frac{\varepsilon^{0\mu\sigma\nu}}{\sqrt{g}} \left( s_{,\tau}\gamma_{(r),\sigma} \lambda_{(r),\nu} + s_{,\nu} \gamma_{(r),\tau}\lambda_{(r),\sigma} + s_{,\sigma}\gamma_{(r),\nu} \lambda_{(r),\tau} \right), \nonumber \\
&= -\left( \rho q \frac{\partial{\cal G}'}{\partial q} \right)_{,\tau} + (\rho q)_{,\tau} \frac{\partial{\cal G}'}{\partial q} + \left(\frac{\partial{\cal G}'}{\partial q} \sum_{r=1}^N \frac{\varepsilon^{0\mu\sigma\nu}}{\sqrt{g}} \left( s_{,\tau}\gamma_{(r),\sigma} \lambda_{(r),\nu} + s_{,\nu} \gamma_{(r),\tau}\lambda_{(r),\sigma} + s_{,\sigma}\gamma_{(r),\nu} \lambda_{(r),\tau} \right) \right)_{:\mu} \nonumber \\
&\quad\, -\frac{\partial{\cal G}'}{\partial q} \sum_{r=1}^N \frac{\varepsilon^{0\mu\sigma\nu}}{\sqrt{g}} \left( s_{,\mu:\tau}\gamma_{(r),\sigma} \lambda_{(r),\nu} + s_{,\nu} \gamma_{(r),\mu:\tau}\lambda_{(r),\sigma} + s_{,\sigma}\gamma_{(r),\nu} \lambda_{(r),\mu:\tau} \right), \nonumber \\
&= -\left( \rho q \frac{\partial{\cal G}'}{\partial q} \right)_{,\tau} + (\rho q)_{,\tau} \frac{\partial{\cal G}'}{\partial q} + \left(\frac{\partial{\cal G}'}{\partial q} \sum_{r=1}^N \frac{\varepsilon^{0\mu\sigma\nu}}{\sqrt{g}} \left( s_{,\tau}\gamma_{(r),\sigma} \lambda_{(r),\nu} + s_{,\nu} \gamma_{(r),\tau}\lambda_{(r),\sigma} + s_{,\sigma}\gamma_{(r),\nu} \lambda_{(r),\tau} \right) \right)_{:\mu} \nonumber \\
&\quad\, -\frac{\partial{\cal G}'}{\partial q} (\rho q)_{,\tau}, \nonumber \\
&= -\left(\rho\frac{\partial{\cal G}'}{\partial q} \left[q\,\delta^\mu_\tau-\sum_{r=1}^N \frac{\varepsilon^{0\mu\sigma\nu}}{\sqrt{g}\rho} \left( s_{,\tau}\gamma_{(r),\sigma} \lambda_{(r),\nu} + s_{,\nu} \gamma_{(r),\tau}\lambda_{(r),\sigma} + s_{,\sigma}\gamma_{(r),\nu} \lambda_{(r),\tau} \right) \right]\right)_{:\mu}.
\end{align}
Since $\bar\delta v_i$, $\bar\delta \rho$, $\bar\delta s$ and $h^{0\mu}$ are zero, (2.5) in \citet{Charron18b} leads to $\bar\delta{\cal L}=-\rho g^{0\tau} \bar\delta v_\tau={J^\mu}_{:\mu}$, where
\begin{align}
J^\mu = \epsilon \rho\frac{\partial{\cal G}}{\partial q} \left[g^{0\mu}q-\sum_{r=1}^N \frac{\varepsilon^{0\mu\sigma\nu}}{\sqrt{g}\rho} \left( s^{:0}\gamma_{(r),\sigma} \lambda_{(r),\nu} + s_{,\nu} {\gamma_{(r)}}^{:0}\lambda_{(r),\sigma} + s_{,\sigma}\gamma_{(r),\nu} {\lambda_{(r)}}^{:0} \right) \right].
\end{align}
\item The 4-vector $J^\mu-\epsilon\rho u^\mu q\,\partial{\cal G}/\partial q$ may be written
\begin{align}
J^\mu - \epsilon\rho u^\mu q \frac{\partial{\cal G}}{\partial q} &= \epsilon\rho \frac{\partial{\cal G}}{\partial q} \left[ (g^{0\mu}-u^\mu)q - \sum_{r=1}^N \frac{\varepsilon^{0\mu\sigma\nu}}{\sqrt{g}\rho} \left( s^{:0}\gamma_{(r),\sigma} \lambda_{(r),\nu} + s_{,\nu} {\gamma_{(r)}}^{:0}\lambda_{(r),\sigma} + s_{,\sigma}\gamma_{(r),\nu} {\lambda_{(r)}}^{:0} \right)\right], \nonumber \\
&=-\epsilon\rho \frac{\partial{\cal G}}{\partial q} \left[ h^{\mu\alpha} v_\alpha\,q + g^{0\tau}\sum_{r=1}^N \frac{\varepsilon^{0\mu\sigma\nu}}{\sqrt{g}\rho} \left( s_{,\tau}\gamma_{(r),\sigma} \lambda_{(r),\nu} + s_{,\nu} {\gamma_{(r),\tau}}\lambda_{(r),\sigma} + s_{,\sigma}\gamma_{(r),\nu} {\lambda_{(r),\tau}} \right)\right], \nonumber \\
&=-\epsilon\rho \frac{\partial{\cal G}}{\partial q} \Bigg[ h^{\mu\alpha} v_\alpha\,q \, + \nonumber \\
&\qquad\qquad\qquad (u^\tau-h^{\tau\alpha}v_\alpha)\sum_{r=1}^N \frac{\varepsilon^{0\mu\sigma\nu}}{\sqrt{g}\rho} \left( s_{,\tau}\gamma_{(r),\sigma} \lambda_{(r),\nu} + s_{,\nu} {\gamma_{(r),\tau}}\lambda_{(r),\sigma} + s_{,\sigma}\gamma_{(r),\nu} {\lambda_{(r),\tau}} \right)\Bigg], \nonumber \\
&=-\epsilon\rho \frac{\partial{\cal G}}{\partial q} v_\alpha \left[ h^{\mu\alpha} q - h^{\tau\alpha}\sum_{r=1}^N \frac{\varepsilon^{0\mu\sigma\nu}}{\sqrt{g}\rho} \left( s_{,\tau}\gamma_{(r),\sigma} \lambda_{(r),\nu} + s_{,\nu} {\gamma_{(r),\tau}}\lambda_{(r),\sigma} + s_{,\sigma}\gamma_{(r),\nu} {\lambda_{(r),\tau}} \right)\right] \nonumber \\
&\quad\, -\epsilon\frac{\partial{\cal G}}{\partial q} \sum_{r=1}^N \frac{\varepsilon^{0\mu\sigma\nu}}{\sqrt{g}} \left( \frac{ds}{dt}\gamma_{(r),\sigma} \lambda_{(r),\nu} + s_{,\nu} \frac{d\gamma_{(r)}}{dt}\lambda_{(r),\sigma} + s_{,\sigma}\gamma_{(r),\nu} \frac{d\lambda_{(r)}}{dt} \right).
\end{align}
However,
\begin{align}
h^{\mu\alpha} q - h^{\tau\alpha}\sum_{r=1}^N \frac{\varepsilon^{0\mu\sigma\nu}}{\sqrt{g}\rho} \left( s_{,\tau}\gamma_{(r),\sigma} \lambda_{(r),\nu} + s_{,\nu} {\gamma_{(r),\tau}}\lambda_{(r),\sigma} + s_{,\sigma}\gamma_{(r),\nu} {\lambda_{(r),\tau}} \right) \equiv 0
\end{align}
identically, while $ds/dt$, $d\gamma_{(r)}/dt$ and $d\lambda_{(r)}/dt$ vanish on-shell. Therefore, $J^\mu=\epsilon\rho u^\mu q\,\partial{\cal G}/\partial q$ on-shell.
\item Consider the functional
\begin{align}
C=\int_{\cal D} d^3x \sqrt{g} \rho q {\cal A}(s)=\sum_{r=1}^N\int_{\cal D} d^3x \,\varepsilon^{0ijk} s_{,i} \gamma_{(r),j} \lambda_{(r),k} {\cal A}(s). \label{naip}
\end{align}
Define ${\cal K}(s)$ as the primitive of $\cal A$:
\begin{align}
{\cal K}(s) \equiv {\cal K}(s_0) + \int_{s_0}^s ds' {\cal A}(s')
\end{align}
such that ${\cal A}\equiv d{\cal K}/ds$, with $s_0$ a constant. One may then rewrite \eqref{naip} as
\begin{align}
C&=\sum_{r=1}^N\int_{\cal D} d^3x \,\varepsilon^{0ijk} \gamma_{(r),j} \lambda_{(r),k} {\cal K}(s)_{,i}, \nonumber \\
&=\sum_{r=1}^N\int_{\cal D} d^3x \left( \varepsilon^{0ijk} \gamma_{(r),j} \lambda_{(r),k} {\cal K}(s) \right)_{,i}, \nonumber \\
&=\sum_{r=1}^N\oint_{\partial{\cal D}} dS_i\,\left( \frac{\varepsilon^{0ijk}}{\sqrt{g}} \gamma_{(r),j} \lambda_{(r),k} {\cal K}(s) \right).
\end{align}
This functional $C$ being a boundary term, its first variation $\delta C$ vanishes and $C$ is therefore a trivial Casimir invariant.
\end{enumerate}
\end{appendices}

\bibliographystyle{ametsoc2014}
\bibliography{refer}

\end{document}